\documentclass{aa}

\usepackage{graphicx}
\usepackage{txfonts}
\usepackage{natbib}
\bibpunct{(}{)}{;}{a}{}{,}

\begin{document}

\title{Empirical Color Transformations Between SDSS Photometry and
Other Photometric Systems} 

\author{K.\ Jordi{\thanks{Guest User, Canadian Astronomy Data Centre,
which is operated by the Dominion Astrophysical Observatory for the
National Research Council of Canada's Herzberg Institute of
Astrophysics.}} \and E.K.\ Grebel \and K.\ Ammon}

\institute{Astronomical Institute of the University of Basel, 
Department of Physics and Astronomy, Venusstrasse 7, CH-4102 Binningen,
Switzerland\\
           \email{jordi@astro.unibas.ch, grebel@astro.unibas.ch, ammon@astro.unibas.ch}}

\date{Received August 14, 2006; accepted August 28, 2006} 

\abstract 
{} 
{We present \emph{empirical} color transformations between the Sloan
Digital Sky Survey (SDSS) \emph{ugriz} photometry and Johnson-Cousins
\emph{UBVRI} system and Becker's \emph{RGU} system, respectively.
Owing to the magnitude of data that is becoming available in the SDSS
photometric system it is particularly important to be able to convert
between this new system and traditional photometric systems. Unlike
earlier published transformations we based our calculations on stars
actually measured by the SDSS with the SDSS 2.5-m telescope.  The
photometric database of the SDSS provides in a sense a single-epoch
set of 'tertiary standards' covering more than one quarter of the sky.
Our transformations should facilitate their use to easily and reliably
derive the corresponding approximate Johnson-Cousins or \emph{RGU}
magnitudes.  }
{The SDSS survey covers a number of areas that were previously
established as standard fields in the Johnson-Cousins system, in
particular, fields established by Landolt and by Stetson. We used
these overlapping fields to create well-photometered star samples on
which our calculated transformations are based.  For the \emph{RGU}
photometry we used fields observed in the framework of the new Basel
high-latitude field star survey.}  
{We calculated \emph{empirical} color transformations between SDSS
photometry and Johnson-Cousins \emph{UBVRI} and Becker's \emph{RGU} system.  For all
transformations we found linear relations to be sufficient.
Furthermore we showed that the transformations between the
Johnson-Cousins and the SDSS system have a slight dependence on
metallicity.}
{}


\maketitle

\section{Introduction}

The Sloan Digital Sky Survey (SDSS) is the largest photometric and 
spectroscopic survey in the optical wavelength range.  The SDSS is 
mapping one quarter of the entire sky and is measuring positions and 
magnitudes for over 100 million celestial objects \citep{York2000, 
Grebel2001}. The SDSS photometric system was specifically
designed for the survey \citep{Fukugita1996, Smith2002}, but many
observatories worldwide have since purchased SDSS filters as well.
The five SDSS \emph{ugriz} filters  are a modified Thuan-Gunn
system~\citep{Thuan1976}.  Their most prominent feature are the wide
passbands, which cover an effective wavelength range of $\sim 380$ nm
to $\sim 920$ nm, and the lack of overlap between neighboring
passbands.  These properties ensure a high efficiency for faint object
measurements, independent spectral information in each band, and
coverage of essentially the entire optical wavelength range accessible
from the ground. 

The SDSS provides a photometric catalog of unprecedented depth,
homogeneity, and quality. Owing to the magnitude of the data that are
becoming publicly available in this new photometric system
\citep{Stoughton2002, Abazajian2003, Abazajian2004, Abazajian2005,
Adelman-McCarthy2006}, it is important to have well-calibrated
transformation relations between this system and traditional
photometric systems such as the Johnson-Cousins system
\citep{Johnson1953, Cousins1976}.  It is easy to imagine situations
where one wishes to know, e.g., the V-band magnitude of a star of
interest that happens to be in the SDSS database and for which no
other photometry is available. Moreover, the large area coverage of
the SDSS and the high quality of its drift-scan photometry make it
also suitable as a source of \textquoteleft tertiary
standards\textquoteright, although there is no information on
variability for the majority of the SDSS objects.  This disadvantage
is compensated by the large number of photometered sources even within
a small patch of the sky.  Thus a few variable objects will merely
appear as outliers and will not have a major effect on a photometric
transformation.  Also, if observations in a traditional photometric
system are being obtained during a non-photometric night, existing
transformable SDSS photometry of the same field will prove very useful
for at least an approximate photometric calibration \citep{Koch2004a,
Koch2004b}.  Other large-area photometric survey catalogs are already
being used in this manner; for instance, \citet{Udalski1998,
Udalski2000, Udalski2002, Epchtein1999, Zaritsky2002, Zaritsky2004,
Skrutskie2006}.   

Prior to the start of SDSS observations, \citet{Fukugita1996} derived
theoretical transformation relations between the Johnson-Cousins
system and the SDSS \emph{u$'$g$'$r$'$i$'$z$'$} system.  The primes
refer to the filter-detector combination envisioned to be used at the
$20''$ photometric monitoring telescope at Apache Point Observatory.
This auxiliary telescope observes SDSS photometric standards while the
science observations are done in drift-scan mode with the actual SDSS
survey telescope, a dedicated 2.5-m telescope at the
same site (see, e.g., \citet{York2000, Stoughton2002, Gunn2006}).
\citet{Fukugita1996} calculated synthetic magnitudes using the planned
filter-detector combinations and spectral energy distributions of
stars from the \citet{Gunn1983} spectrophotometric atlas.  These
synthetic magnitudes were then used to determine photometric
transformations.

\citet{Smith2002} calculated transformations between the SDSS
\emph{u$'$g$'$r$'$i$'$z$'$} system and the Johnson--Cousins
photometric system based on actual observations in
\emph{u$'$g$'$r$'$i$'$z$'$} filters.  In this context the primes refer
to the SDSS filter-detector combination used at the 1.0-m telescope at
the US Naval Observatory (USNO), Flagstaff Station.  These observations were
used to set up a system of 158 bright standard stars that define the
\emph{u$'$g$'$r$'$i$'$z$'$} system and to derive the above mentioned
transformation equations.  \citet{Smith2002} point out that there are
``small but significant'' differences between the USNO SDSS filters
and the SDSS filters used at the 2.2-m telescope at Apache Point
Observatory, leading to expected systematic differences between the
USNO magnitudes and the final SDSS magnitudes.  

Similarly, \citet{Karaali2005} used observations obtained in
\emph{u$'$g$'$r$'$} filters at the Isaac Newton Telescope (INT) at La Palma,
Spain.  These filters were designed to reproduce the SDSS system.
These data were complemented by \citet{Landolt1992} \emph{UBV}
standard star photometry and used to calculate transformations between
the INT SDSS \emph{u$'$g$'$r$'$} filter-detector combination and
standard Johnson--Cousins photometry.  \citet{Karaali2005} presented
for the first time transformation equations depending on two colors.
Since the INT filters and detector differ from the actual
filter-detector combinations used by the SDSS, again systematic
deviations are to be expected.

Direct empirical transformations between SDSS point-source photometry
from the ``early data release'' (EDR; \citet{Stoughton2002}) and
certain Johnson-Cousins filters were calculated by
\citet{Odenkirchen2001} and by \citet{Rave2003}.  \citet{Rave2003}
based their transformations on resolved stellar photometry of the
Draco dwarf spheroidal galaxy obtained with the SDSS as well as with
various other telescopes in different variants of the Johnson-Cousins
system.  The SDSS EDR photometry was then superseded by later SDSS
data releases \citep{Abazajian2003, Abazajian2004, Abazajian2005,
Adelman-McCarthy2006}.  

\begin{figure}
\centering
\includegraphics[width=88mm]{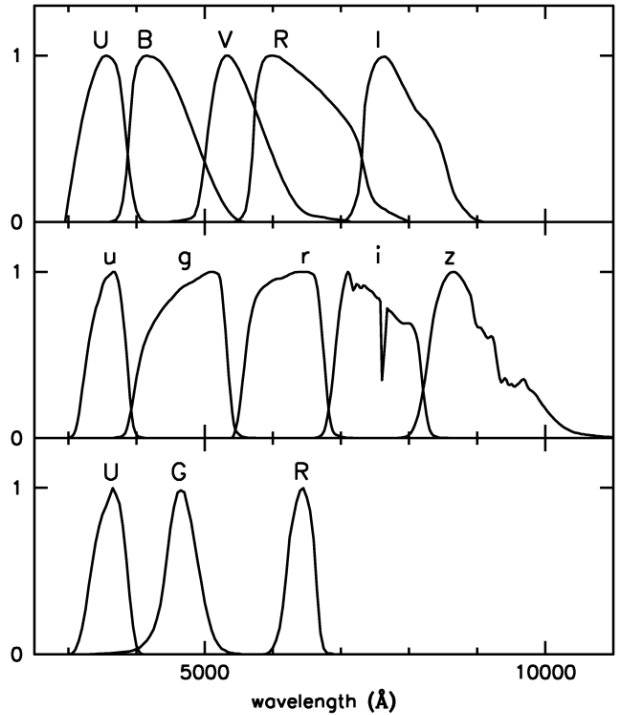}
\caption{Comparison of the (normalized) filter curves of the
Johnson-Cousins \emph{UBVRI} passbands (upper panel, the SDSS
\emph{ugriz} passbands (middle panel), and Becker's \emph{RGU}
photometry in the lower panel.}
\label{fig:filters}
\end{figure}

A recent \emph{theoretical} transformation was carried out by
\citet{Girardi2004}.  These authors used the actual SDSS passbands and
CCD sensitivity curves of the SDSS camera \citep{Gunn1998} employed at
the 2.5-m SDSS telescope at Apache Point Observatory  to transform
theoretical stellar isochrones into the SDSS \emph{ugriz} system.

The third photometric system that we are considering here is the
\emph{RGU} system by Becker \citep{Becker1946a}, a broad-band
photometric system that was initially introduced with the goal of
measuring stellar parameters and Galactic structure.  No earlier
transformations between SDSS photometry and \emph{RGU} photometry have
been published. For transformation relations between \emph{RGU} and \emph{UBV} see \citet{Steinlin1968a, Steinlin1968b, Bell1972, Buser1978a, Buser1978b}. While \emph{RGU} photometry is not widely used, it
plays a continuing role due to the existence of the ``new Basel
high-latitude field star survey'' \citep{Buser1985, Buser1998}.  The
goal of this single-epoch photographic survey is to map the Galactic
density structure and the metallicity distribution in the various
components of the Milky Way.  For this purpose, 14 intermediate and
high-latitude fields were observed, in which three-color \emph{RGU}
photometry of over 18\,000 stars was measured. The limiting magnitude
of this Galactic survey is approximately $G\sim19$ or $V\sim18$.  This
survey may seem to be obsolescent by modern standards due to its small
number of stars, the shallow exposures, the limited photographic
resolution, and the availability of much deeper, more homogeneous, and
much larger CCD surveys.  On the other hand, this photographic survey
has played an important role in uncovering the structure of our
Galaxy (e.g., \cite{Buser1999}). Transformation relations between SDSS and \emph{RGU}
photometry help to evaluate the earlier findings with respect to
modern explorations of the structure of the Milky Way (e.g.,
\citet{Chen2001, Juric2005}). The Basel survey provides star counts (or space densities) also for distances relatively close to the Sun, whereas there is a
gap at these distances in studies carried out using SDSS data due
to the saturation limit at $r \sim 14$.  Space densities for stars
with brighter apparent magnitudes can be combined with Hipparchos
parallaxes, thus providing reliable Galactic model parameters. Moreover, the metallicity sensitivity
of the \emph{RGU} system (e.g., \cite{Buser1990}) adds continuing
value to the Basel high-latitude field star survey.  Finally, this
survey comprises a number of important Galactic sight lines that are
not included in the currently available wide-field CCD surveys, and
the analysis of these sight lines is continuing.

Here we present \emph{empirical} color transformations between the
SDSS \emph{ugriz} photometric system using point-source photometry
obtained with the SDSS 2.5-m telescope at Apache Point Observatory,
and the Johnson-Cousins \emph{UBVRI} system as well as the \emph{RGU}
system, respectively. These empirical colors have the advantage of not
having to rely on synthetic colors or on passband-detector
combinations of other telescopes.  Hence, they provide the most direct
transformations possible.  In Sect. 2 we present the sources of our
star samples and the criteria applied to choose the stars. Section 3
contains our calculations and the resulting transformation equations.
In Sect. 4 we discuss our results and some possible future
developments.

\section{Data}

Our goal is to derive empirical transformations between the SDSS
photometry as defined by the public SDSS data and the two other aforementioned
photometric systems while minimizing the dependence on specific
filter-detector combinations that deviate from the generic ones.  The
data used for our transformations were taken from the four different
sources described below. 

The SDSS data release 4 (SDSS DR4; \citet{Adelman-McCarthy2006})
comprises stellar point-source magnitudes provided by the SDSS
photometry pipelines \citep{Lupton2001, Stoughton2002}.  This
point-source photometry is not expected to change anymore between DR4
and later releases (e.g., \citet{Adelman-McCarthy2006}).

\subsection{Johnson-Cousins photometry}

For Johnson-Cousins photometry the most generic standard star database
available that lends itself to a comparison is the system of standard
stars set up by A.\ Landolt, in particular the catalog published by
\citet{Landolt1992}.  This catalog contains an extensive set of mainly
equatorial fields observed repeatedly with photoelectric aperture
photometry.  A number of the Landolt fields overlap with the area
scanned by the SDSS.  The Landolt photometry is widely used to
transform and calibrate imaging data.  The Landolt stars cover a
$V$-magnitude range of approximately 11.5 to 16.0.  With respect to
SDSS photometry, the Landolt standard stars have one drawback:  Many
of them are brighter than the saturation cutoff in the SDSS system ($
r \approx 14$).  Nonetheless, the Landolt stars are valuable
particularly for the transformation of the $U$-band photometry as we
will see later. 

The problem of the reduced number of stars in common between the
Landolt catalog and the SDSS due to saturation is alleviated by the
extension of Landolt's standard stars to fainter magnitudes by
\citet{Stetson2000}.  Stetson used a large set of multi-epoch CCD
observations centered on Landolt fields and other regions in the sky
and reduced them in a homogeneous manner tied to the Landolt
\emph{UBVRI} system.  The larger area coverage and greater sensitivity
of the CCD observations as compared to the earlier photomultiplier
observations permitted Stetson to include stars down to $V \approx 19$
or 20.  Since 2000, Stetson has been publishing a gradually growing
list of suitable faint stars \citep{Stetson2000} with repeat
observations at the website of the Canadian Astronomy Data
Center\footnote{http://cadcwww.dao.nrc.ca/standards/}.  The Stetson
catalog contains only stars that were observed at least five times
under photometric conditions and the standard error of the mean
magnitude is less than 0.02 mag in at least two of the four filters.
Stetson's data base also contains fields not covered by Landolt, e.g.,
fields in globular clusters and in nearby resolved dwarf galaxies.
While Landolt's original fields contained mainly Population I stars,
Stetson's new fields also include a sizable fraction of Population II
stars.  

\begin{table}
\begin{minipage}[t]{\columnwidth}
\caption{The 23 Stetson fields that overlap with the SDSS DR4 sky
coverage, and the number of matched stars.}
\label{tab:stetson}
\centering
\renewcommand{\footnoterule}{}
\begin{tabular}{lrr}
\hline
\hline
Fields & Stetson Stars & Matches\footnote{with \emph{clean}
photometry\\ }\\
\hline
L92 & 214 & 138\\
L95 & 427 & 250\\
NGC\,2419 & 1\,188 & 520\\
NGC\,2420 & 188 & 54\\
NGC\,2683 & 27 & 7\\
PG\,0918 & 122 & 53\\
L101 & 118 & 67\\
Leo\,I &1\,840 & 508\\
PG\,1047 & 67  & 36\\
NGC\,5194 & 39 & 3\\
NGC\,5272 & 432& 111\\
NGC\,5466 & 29 &3\\
L106\_550 & 16 & 5\\
Pal\,5 & 65 & 48\\
L107 & 729 & 490\\
Pal\,14 &163 & 116\\
Draco & 529 & 256\\
L112 & 74 & 26\\
NGC\,7078 & 967  & 114\\
NGC\,7089 & 377 & 26\\
L113 & 484 & 320\\
PG\,2213 & 36 & 16\\
Pegasus & 38 & 28\\
\textbf{Total} & \textbf{8\,169}  & \textbf{3\,195}\\
\hline
\end{tabular}
\end{minipage}
\end{table}

For our work the Stetson fields published as of January 2005 were
used. As most of the SDSS DR4 sky coverage is in the northern part of
the sky, it was straightforward to select those fields from the
Stetson catalog that overlap with the SDSS DR4:  30 fields were
identified and all stars available from the SDSS DR4 database within
these fields were downloaded. For the subsequent matching and
calculation of transformation relations only SDSS stars with
\emph{clean} photometry were used. The combination of flags describing
the \emph{clean} photometry can be found on the SDSS DR4
webpage\footnote{http://cas.sdss.org/astrodr4/en/help/docs/sql\_help.asp\#clean}.
This flag combination excludes stars whose photometry may be
questionable for a number of reasons, e.g., due to saturation,
overlaps with other objects (blends), location at the edge of a frame,
large errors in fitting a point spread function, etc.  

We then matched the Stetson stars and the SDSS stars by their
coordinates.  The coordinates of the stars in DR4 are measured with an
accuracy of less than $0.1''$ rms per coordinate
\citep{Adelman-McCarthy2006}. In the Stetson catalog the coordinates
are published with an accuracy of $0.15''$ for the right ascension and
$0.1''$ for the declination.  Two stars -- one from each sample -- are
considered equal if their angular separation is smaller than $0.5''$:
$\sqrt{(\alpha_1-\alpha_2)^2cos(\delta_1)^2+(\delta_1-\delta_2)^2}\leq0.5''$,
where $\alpha$ and $\delta$ stand for the right ascension and
declination of the stars in arcsec.  

\begin{figure}
\centering
\includegraphics[width=88mm]{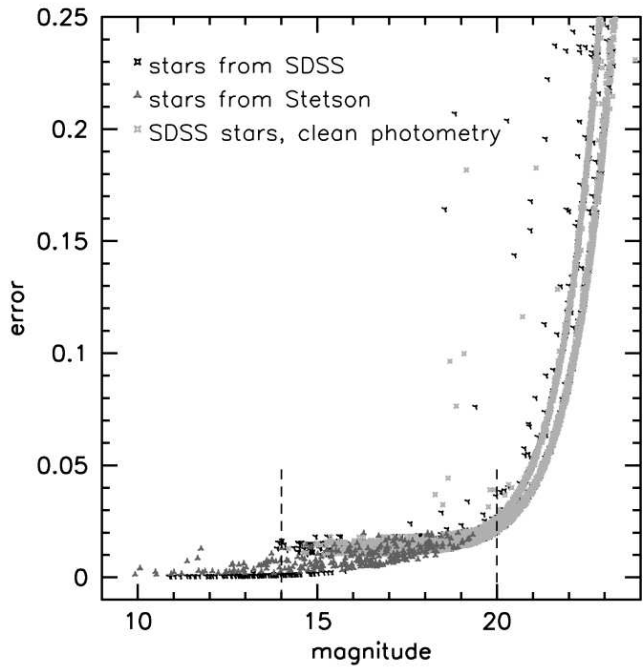}
\caption{For field L95 we plot the $R$ magnitude of Stetson's standard
stars (dark gray triangles) and the $r$ magnitude of the SDSS stars
(black crosses) against their photometric errors. The light gray
crosses are those SDSS stars with \emph{clean} photometry. The vertical dashed lines indicate the magnitude intervall in which possible matches in field L95 will lie.} 
\label{error}
\end{figure}

Although the matching radius was chosen generously we do not find a
matching SDSS star for each Stetson star (see
Table~\ref{tab:stetson}). The \emph{clean} photometry rule removes
between 30 and 40 percent of our initial star sample.  Furthermore,
the SDSS photometric pipeline only detects a reduced number of stars
in crowded fields.  This affects in particular fields in globular
clusters such as NGC 2419, NGC 5272, NGC 5466, NGC 7078, and NGC 7089,
and fields with luminous extended galaxies such as NGC 2683.  For
fields with a very high degree of crowding, no SDSS data are available
since these fields were intentionally omitted during the pipeline
reduction process.  Owing to these limitations, our original number of
30 common fields is actually reduced to 23 fields in which common
stars in Stetson's and the SDSS catalogs could be identified.  Fields
without \emph{any} common stellar objects with clean photometry
include regions centered on the galaxies M81, NGC 4526, and NGC 4736,
on the globular clusters M5 and M13, and on fields with stars that are
all saturated in the SDSS like L114-750.  In Table~\ref{tab:stetson}
the number of matches for each of the remaining 23 fields is listed.
In Fig.~\ref{error} we plot the $R$ magnitude of Stetson stars
against their photometric error and the $r$ magnitude of SDSS stars
against their photometric error in order to illustrate how many stars
we lose in the matching process.  The
Stetson stars are all brighter than $20^{th}$ magnitude, but at the
same time the brightest SDSS stars with \emph{clean} photometry are of
$14^{th}$ magnitude. Possible matches between stars in the Stetson and in the SDSS database will therefore lie within this magnitude interval (see dashed vertical lines in Fig.~\ref{error}).

Since Stetson's fields do not include $U$-band photometry in the fields
that overlap with the SDSS, we complemented the Stetson data by
standard stars from Landolt~\citep{Landolt1992}, our third data
source. Only 54 stars from the Landolt catalog overlap with the DR4
sky coverage and have \emph{clean} photometry in the SDSS, hence
transformations involving $U$ are necessarily based on a small subsample
of stars.

\subsection{Becker's \emph{RGU} system}

Our fourth source of data is the ``new Basel high-latitude field star
survey'' \citep{Buser1985, Buser1998}.  The fourteen fields of the
Basel high-latitude field star survey are all specified by a pair of
galactic coordinates \citep{Buser1998}.  These coordinates are not
consistently defined and do not necessarily mark the center of the
field. In most of the fields, they mark one of the corners.  Three 
fields completely overlap with the SDSS DR4 sky coverage and
six others have a partial overlap. The other five fields do not lie
in the region of the sky that the SDSS DR4 has scanned.  For deriving
transformations, those two of the completely overlapping fields were
used that had the best photometric quality.  Table~\ref{tab:basel}
lists the two fields and the number of found matches.  The sample for
the second set of transformations contained a total of 775 stars.

The Basel survey does not list coordinates of the stars in its fields,
so they were matched visually using the Basel finding charts.  In
these prints of the photographic plates each measured star is marked
with its catalog number, making the identification of stars in common
with the SDSS a rather tedious enterprise.  Not for all Basel stars a
matching partner was found.  The SDSS CCD photometry has higher
resolution than the photographic plates of the Basel survey.  In some
cases, the SDSS recognized an object as a galaxy, whereas in the Basel
survey it was treated as a star. In other cases, there was simply no
star in the SDSS where Basel detected one.  Occasionally  the SDSS
detected a fainter object within a radius of $1''$ to $3''$ of the 
dominant star,  whereas in the Basel catalog only one source was
detected.  In all those cases these stars were deleted from our sample
in order to ensure that only reliable photographic photometry was
included.  Moreover, we only used SDSS sources with \emph{clean}
photometry.

\begin{table}
\begin{minipage}[t]{\columnwidth}
\caption{The two fields from the new Basel Catalog that overlap with the SDSS DR4 sky coverage, and the number of matched stars.}
\label{tab:basel}
\centering
\renewcommand{\footnoterule}{}
\begin{tabular}{lrrr}
\hline
\hline
Fields & Stars in Basel & Matches\footnote{with \emph{clean} photometry}\\
\hline
SA94 & 1\,239 & 545\\
SA107&  532 & 230\\
\textbf{Total}& \textbf{1\,771} & \textbf{775}\\
\hline
\end{tabular}
\end{minipage}
\end{table}

\section{Results}

\subsection{Transformations between SDSS and Johnson--Cousins Photometry}

The transformation between the Johnson--Cousins \emph{UBVRI}
photometry system and the SDSS \emph{ugriz} system was carried out
using the following eight general equations:
\begin{eqnarray}
g-V & = & a_1\,(B-V)+b_1\\
r-i & = & a_2\,(R-I)+b_2\\
r-z & = & a_3\,(R-I)+b_3\\
r-R & = & a_4\,(V-R)+b_4\\
u-g & = & a_5\,(U-B)+b_5\,(B-V)+c_5\\
g-B & = & a_6\,(B-V)+b_6\\
g-r & = & a_7\,(V-R)+b_7\\
i-I & = & a_8\,(R-I)+b_8
\end{eqnarray}

Not for all of the standard stars in Stetson's catalog measurements in
all four bands (\emph{BVRI}) are available, hence the number of stars
used differed somewhat depending on the filters used in the
transformation equation. We calculate in all but one case
transformations depending on one color only. For the $u-g$ equation two
colors are used since the SDSS \emph{g} passband overlaps with the
Johnson \emph{B} and \emph{V} passband, so a dependence on $(B-V)$ can
be expected, and more importantly, the variation in position of the
stellar locus owing to temperature, surface gravity, and metallicity 
is particularly
large in this wavelength range (e.g., \citet{GreRob1995, Lenz1998}).

\subsubsection{`Global' transformations between \emph{UBVRI} and
\emph{ugriz}}
\label{globalsection}

In Fig.~\ref{globalstetson} the colors specified on the left-hand side and on the
right-hand side of our transformation equations are plotted against
each other for all the stars used in each of the transformations.  The
resulting linear relations are plotted as solid lines.  With the
exception of the $(r-R)$,$(V-R)$ transformation (equation 4), which
exhibits a pronounced slope change (see also \citet{Fukugita1996}),
the relations are linear to first order.  
In Table~\ref{globalstetson} the resulting coefficients for the
\textquoteleft global\textquoteright~transformations are listed.  

\begin{table*} 
\centering
\caption{Coefficients of the \textquoteleft global\textquoteright\space 
transformations between \emph{UBVRI} and \emph{ugriz} (equations 1--8)} 
\label{globalstetson}
\begin{tabular}{cccl} 
\hline 
\hline 
Color & Color Term & Zeropoint & Range\\ \hline 
$g-V$ & $(0.630\pm0.002)\,(B-V)$ & $-(0.124\pm0.002)$ &\\ 
$r-i$ & $(1.007\pm0.005)\,(R-I)$ & $-(0.236\pm0.003)$ &\\ 
$r-z$ & $(1.584\pm0.008)\,(R-I)$ & $-(0.386\pm0.005)$ &\\ 
$r-R$ & $(0.267\pm0.005)\,(V-R)$ & $+(0.088\pm0.003)$ &$V-R\le
0.93\label{marker3.9a}$\\ 
$r-R$ & $(0.77\pm0.04)\,(V-R)$ & $-(0.37\pm0.04)$ &$V-R>0.93$\\ 
$u-g$ & $(0.750\pm0.050)\,(U-B)+(0.770\pm0.070)\,(B-V)$ &
$+(0.720\pm0.040)$&\\ 
$g-B$ & $-(0.370\pm0.002)\,(B-V)$ & $-(0.124\pm0.002)$ &\\
$g-r$ & $(1.646\pm0.008)\,(V-R)$ & $-(0.139\pm0.004)$ &\\
$i-I$ & $(0.247\pm0.003)\,(R-I)$ & $+(0.329\pm0.002)$ & \\
\hline
\end{tabular} 
\end{table*} 

\begin{figure*}
\centering
\includegraphics[width=16cm]{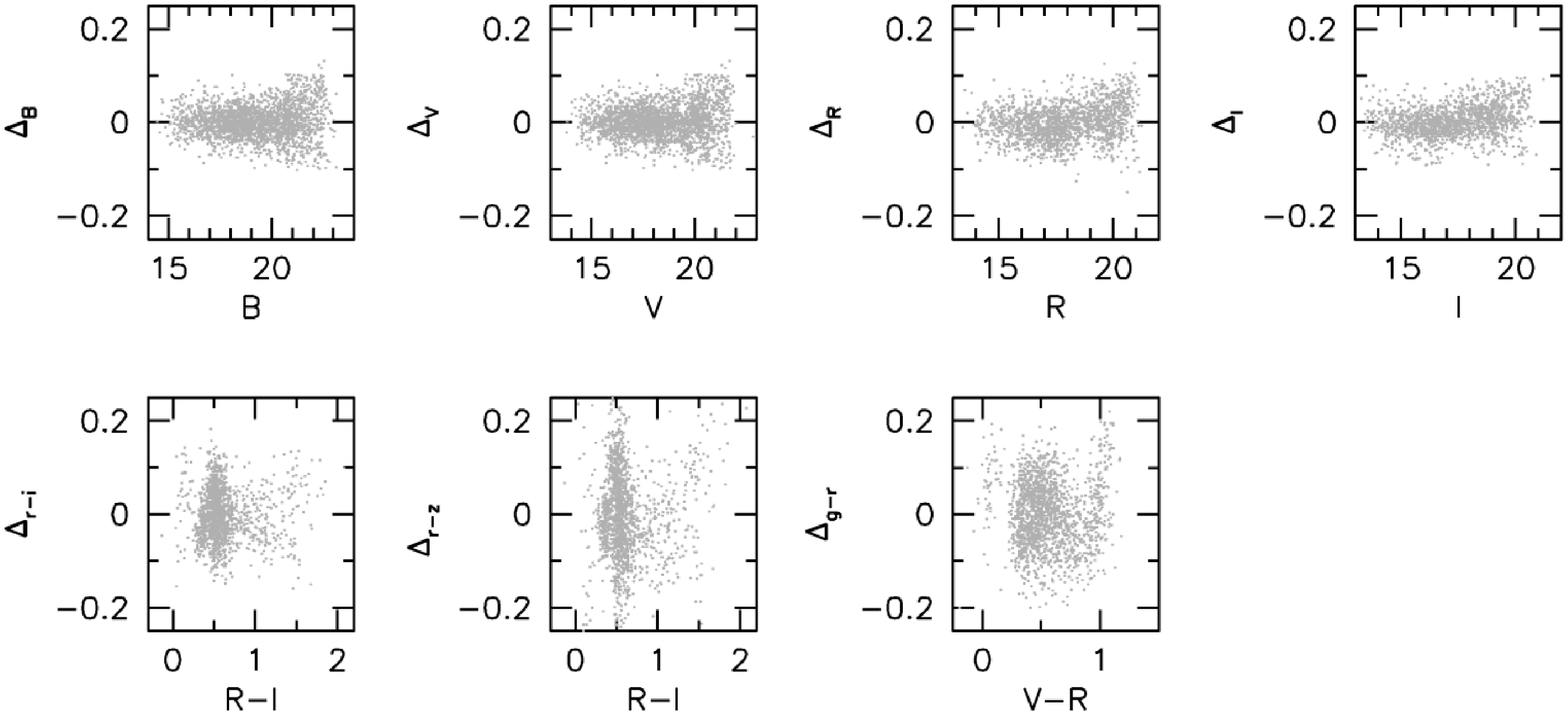}
\caption{The deviations of the measured magnitude and color values from the
calculated values is shown, using the notation $\Delta_{\rm magnitude}
= ({\rm magnitude})_{measured}-({\rm magnitude})_{calculated}$.
For the magnitudes
$B$, $V$, and $R$ the deviations are smaller than for the three colors
$(r-i)$, $(r-z)$, and $(g-r)$.}
\label{abweichung}
\end{figure*}

Here \textquoteleft global\textquoteright~ means that the entire star
sample described above was used for the calculation without
differentiating between potential Population I and Population II stars. 
The coefficients were calculated using least-squares minimization.
The individual stars are weighted according to their formal photometric
error. In Fig.~\ref{abweichung} the deviations between the measured
and the calculated magnitudes and colors are shown. The deviations get bigger for
fainter magnitudes.

\begin{figure*} 
\centering
\includegraphics[width=16cm]{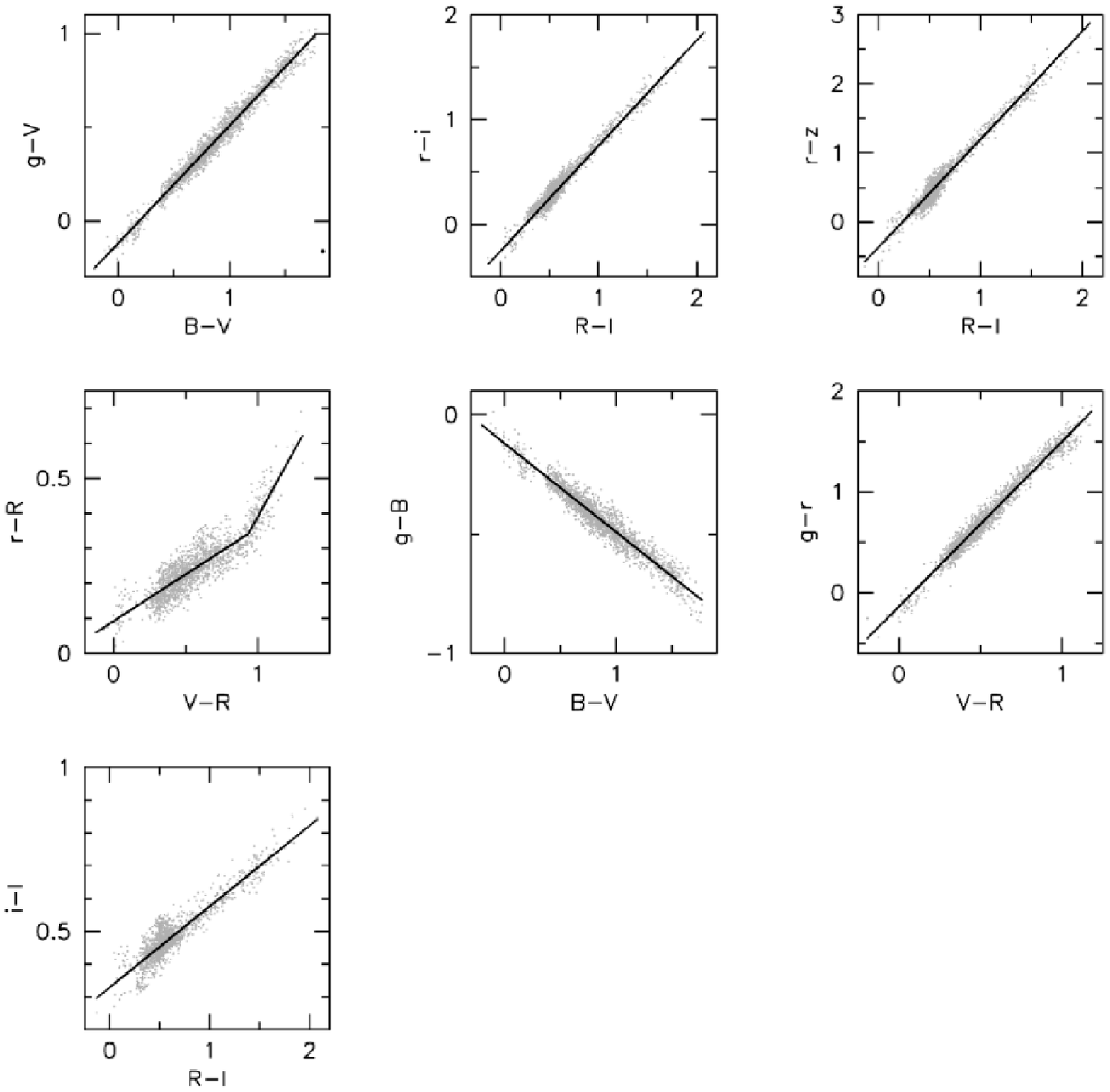}
\caption{The \textquoteleft global\textquoteright\space
transformations between \emph{UBVRI} and \emph{ugriz}. The solid black
line is the best-fit relation.  Its coefficients are listed in
Table~\ref{globalstetson}.} 
\label{global}
\end{figure*} 

\subsubsection{Metallicity-dependent transformations between
\emph{BVRI} and \emph{griz}}

The influence of metallicity on the transformation was
investigated to the extent possible with the available standard star
fields. One of the Stetson fields covers part of the Draco dwarf
spheroidal galaxy. The majority of the stars from this Milky Way
companion has a very low metallicity: Draco's mean metallicity is
[Fe/H] $\sim -2.0$ dex \citep{GGH2003}. Moreover,
Stetson and the SDSS provide photometry for the surroundings of the
similarly metal-poor globular clusters NGC\,2419 ($-2.12$ dex
according to \citet{Harris1996}) and NGC\,7078 ($-2.26$ dex;
\citet{Harris1996}).  Probable member stars of these three objects
were selected based on their color-magnitude diagrams.  This provides
us with a metal-poor sample of ancient Population II stars.  We may
assume that the ages of the three old objects are comparable within
the measurement accuracy (see also \citet{GreGal2004}). 

In contrast, the metallicity of the stars in Landolt's equatorial
fields is not known. We may assume that many of them belong to the
Galactic disk, that Population I stars dominate, and that they have a
range of metallicities. Presumably the metallicity of these stars is
comparatively high and may reach values up to solar.
Our old, metal-poor Population II subsample and our more metal-rich
Population I subsample allow us to empirically assess a possible
impact of metallicity on photometric transformations.  Here our
calculations involve only {\em BVRI} (equations 1 to 4 and 6 to 8).  For equation
5, a distinction between metal-poor and metal-rich stars was not 
possible because of the lack of $U$-band photometry. This is
unfortunate since colors including $U$ or $u$ are particularly
sensitive to stellar parameters including metallicity (see, e.g.,
\citet{GreRob1995, Lenz1998, Helmi2003}).  

\begin{figure*} 
\centering
\includegraphics[width=16cm]{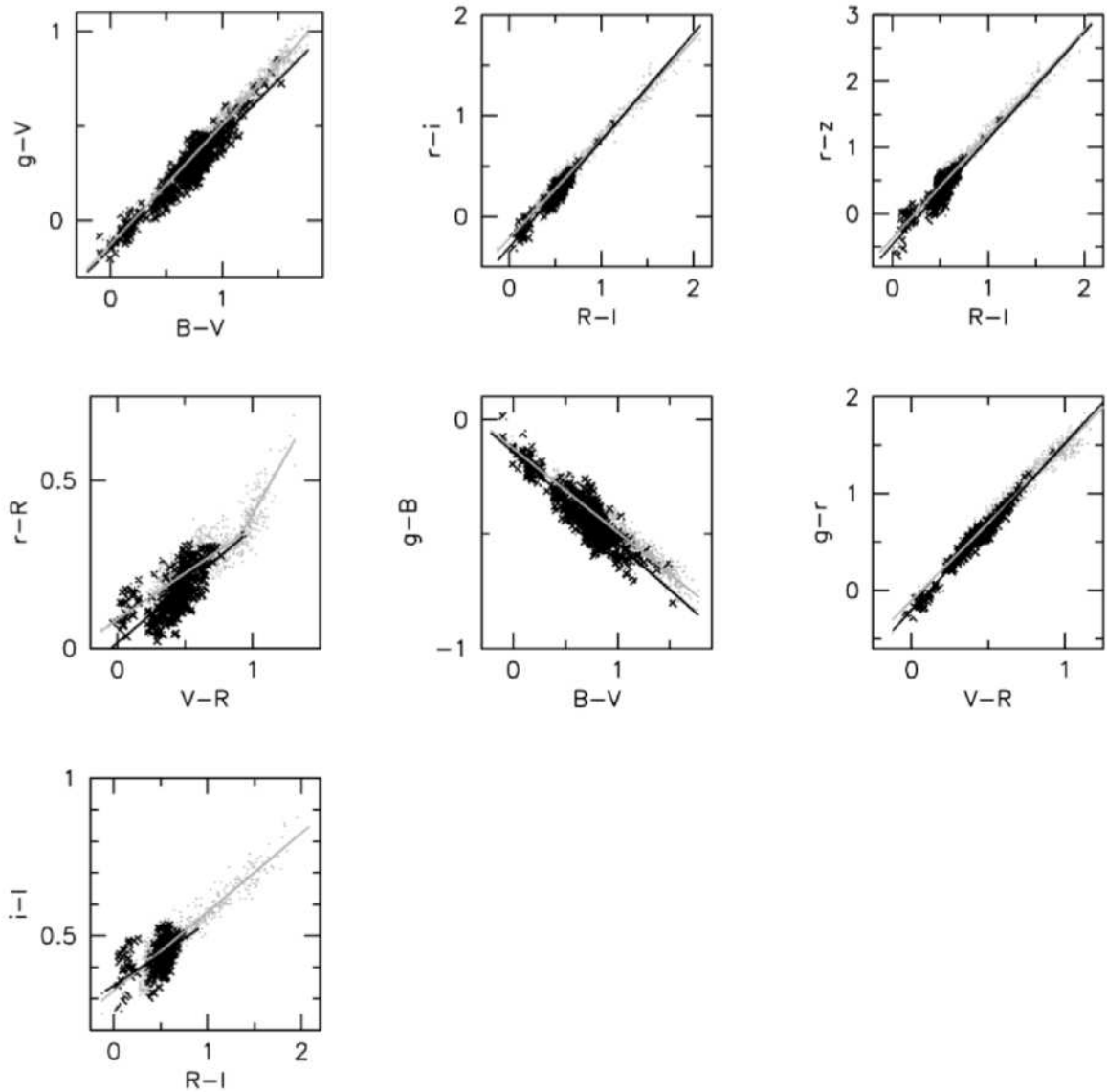}
\caption{The metallicity-dependent transformations. The solid black
line is the fit for metal-poor Population II stars (black dots). The
solid gray line is fit for the Population I stars (gray dots). The
coefficients of the best-fit curves are given in
Table~\ref{metallstetson}.}
\label{metall} 
\end{figure*}

In Table~\ref{metallstetson} the resulting equations are listed. The
resulting linear relations are plotted in Fig.~\ref{metall}.  While
the metal-rich stars are distributed fairly evenly across a wide range
of colors, the metal-poor stars are concentrated within a fairly small
range of colors corresponding mainly to the locus of the red giant
branch.  There are no metal-poor stars in our Population II sample
with colors redder than $(V-R) = 0.93$.  In the diagrams including
$(R-I)$ the locus of the Population I stars is shifted towards
somewhat bluer $(r-i)$ and $(r-z)$ colors as compared to the
Population II stars.  The coefficients of the linear transformation
relations (solid lines in Fig.~\ref{metall}) result in slightly
different slopes for metal-rich and metal-poor stars.  At the main
stellar locus, these deviations are less than the observed scatter in
the colors of the stars in the two samples. Owing to the large scatter in the ($r-R$),($V-R$) and ($i-I$),($R-I$) diagrams, little can be said about possible trends here.

\begin{table*} 
\centering
\caption{Metallicity-dependent transformations between \emph{BVRI}
and \emph{griz} for metal-poor Population II 
and more metal-rich Population I stars. } 
\label{metallstetson} 
\begin{tabular}{clcl} \hline
\hline 
Color & Color Term & Zeropoint & Validity\\ 
\hline 
$g-V$ & $(0.634\pm0.002)\,(B-V)$ & $-(0.127\pm0.002)$ & Population I\\
$g-V$ & $(0.596\pm0.009)\,(B-V)$ & $-(0.148\pm0.007)$ & metal-poor Population II\\ 
$r-i$ & $(0.988\pm0.006)\,(R-I)$ & $-(0.221\pm0.004)$ & Population I\\
$r-i$ & $(1.06\pm0.02)\,(R-I)$ & $-(0.30\pm0.01)$ & metal-poor Population II\\ 
$r-z$ & $(1.568\pm0.009)\,(R-I)$ & $-(0.370\pm0.006)$ & Population I\\ 
$r-z$ & $(1.60\pm0.06)\,(R-I)$ & $-(0.46\pm0.03)$ & metal-poor Population II\\ 
$r-R$ & $(0.275\pm0.006)\,(V-R)$ & $+(0.086\pm0.004)$ &$\mbox{$V-R$}\le0.93; \enskip\mbox{ Population I}$\\ 
$r-R$ & $(0.71\pm0.05)\,(V-R)$  & $-(0.31\pm0.05)$&$\mbox{$V-R$}>0.93; \enskip\mbox{ Population I}$\\
$r-R$ & $(0.34\pm0.02)\,(V-R)$ & $+(0.015\pm0.008)$ &$\mbox{$V-R$}\le0.93; \enskip\mbox{ metal-poor Population II}$\\ 
$g-B$ & $-(0.366\pm0.002)\,(B-V)$& $-(0.126\pm0.002)$ & Population I\\
$g-B$ & $-(0.401\pm0.009)\,(B-V)$ & $-(0.145\pm0.006)$ & metal-poor Population II\\
$g-r$ & $(1.599\pm0.009)\,(V-R)$ & $-(0.106\pm0.006)$ & Population I\\
$g-r$ & $(1.72\pm0.02)\,(V-R) $ & $-(0.198\pm0.007)$ & metal-poor Population II\\
$i-I$ & $(0.251\pm0.003)\,(R-I)$ & $+(0.325\pm0.002)$ & Population I\\
$i-I$ & $(0.21\pm0.02)\,(R-I)$ & $+(0.34\pm0.01)$ & metal-poor Population II\\
\hline 
\end{tabular}
\end{table*} 

\subsubsection{Comparison between our transformations and earlier
work}

\begin{figure*} 
\centering 
\includegraphics[width=16cm]{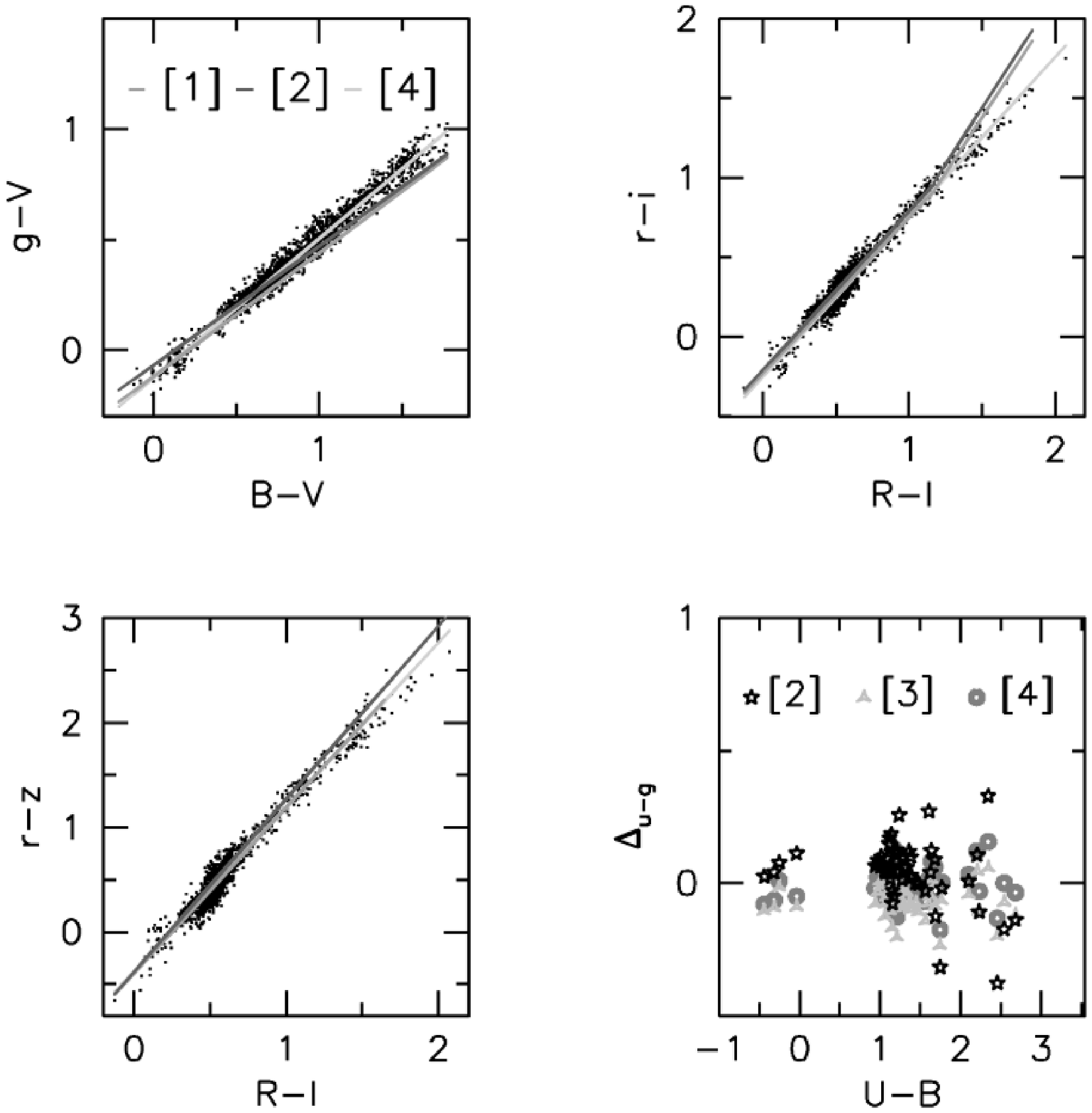}
\caption{Comparison of our transformations [4] with the transformations
by \citet{Fukugita1996} [1], \citet{Smith2002} [2], and
\citet{Karaali2005} [3]. The black dots in the first three panels
represent the entire star sample described in
Section~\ref{globalsection}. The fourth panel shows the deviations between the measured and the calculated $(u-g)$ color.} 
\label{compare} 
\end{figure*}

We compared our \textquoteleft global\textquoteright\, transformations
with transformations published by \citet{Fukugita1996}, \citet{Smith2002} and \citet{Karaali2005}.  \citet{Fukugita1996} used synthetic magnitudes 
from the spectrophotometric atlases of \citet{Gunn1983} and
of \citet{Oke1990} for their transformations. \citet{Smith2002}
presented transformations based on actually measured magnitudes of 158
SDSS standard stars.  Their measurements were not done with the 2.5-m
SDSS telescope, but with the 1.0-m telescope at the US Naval
Observatory, Flagstaff Station.  \citet{Karaali2005} were the first to
do transformations depending on two colors. For their transformations
they used 251 stars of~\citet{Landolt1992} for the \emph{UBV} data and
data from the CASU INT Wide Field Survey measured in filters close to
the SDSS \emph{ugr} passbands. In Fig.~\ref{compare} the color-color
plots show our transformation relations, which are based directly on
the 2.5-m SDSS data, in comparison with the previously published
transformations listed above.

As  Fig.~\ref{compare} demonstrates, for the $(B-V)$ transformation (upper
left panel and equation 1) the \emph{earlier} transformations lie
below our star distribution and have a slightly different slope.
For the $(R-I)$ transformation in equation 2 (upper
right panel) our data do not support a change of slope for red colors
as suggested in the previous studies. The distribution gets broader
for $R-I > 1.15$, and the \citet{Fukugita1996} and \citet{Smith2002}
transformations represent the upper half of the distribution
whereas our fit reproduces the average of the stellar locus.  For equation 3
(lower left panel) the earlier studies and our work show close
agreement.  Transformations for $(u-g)$ (equation 5) were calculated
by \citet{Smith2002}, \citet{Karaali2005}, and by us. Smith et al.'s
equation differs from the others because it only depends on one color.
The lower right diagram in Fig.~\ref{compare} shows the deviation 
between the measured $(u-g)$ color and the calculated $(u-g)$ color. 
The one-color transformations by Smith et al.\ do not reproduce the colors of the combined Landolt and Stetson 
samples very well. The transformation by Karaali et al.\
resembles our result in spite of the somewhat different
filter--telescope combinations.

\subsection{Transformations between SDSS and \emph{RGU} Photometry} 

For the transformation between the \emph{RGU} system and the SDSS
\emph{ugriz} system we defined the following three equations:
\begin{eqnarray} 
U-G & = & a_1\,(u-g)+b_1\label{equationUG}\\ 
G-R & = & a_2\,(g-r)+b_2\\ 
G & = & a_3\,g+b_3 
\end{eqnarray} 
The stars of our SDSS--\emph{RGU} sample show a relatively large 
scatter in the color planes.  One of the reasons for the scatter is the
lower photometric accuracy of the photographic data as compared to
modern CCD data, namely typical internal uncertainties of $\sigma_G =
0.05$, $\sigma_{G-R} = 0.07$, and $\sigma_{U-G} = 0.08$
\citep{Buser1998}.  Some of our 775 common stars lie far off the broad, 
mean stellar loci.  Since the resolution of the photographic data is
relatively low, these deviant
points are most likely false matches or blends in the photographic
photometry.  We thus removed them from our sample before we calculated
the transformation relations. For this set of transformations we used the same procedure as before. The coefficients are listed in Table~\ref{tab:becker}.

\begin{table*} 
\centering
\caption{Coefficients of the transformations between \emph{ugriz} and \emph{RGU} (equations 9--11) } 
\label{tab:becker} 
\begin{tabular}{clc} \hline
\hline 
Color & Color Term & Zeropoint \\ 
\hline 
$U-G$ & $ (0.95\pm0.01)\,(u-g)$ & $+0.16\pm0.02$\\
$G-R$ & $ (1.07\pm0.02)\,(g-r)$ & $+0.64\pm0.02$\\ 
$G$   & $ (0.989\pm0.005)\,g$   & $00.37\pm0.08$\\
\hline
\end{tabular} 
\end{table*}

\begin{figure*} 
\centering
\includegraphics[width=16cm]{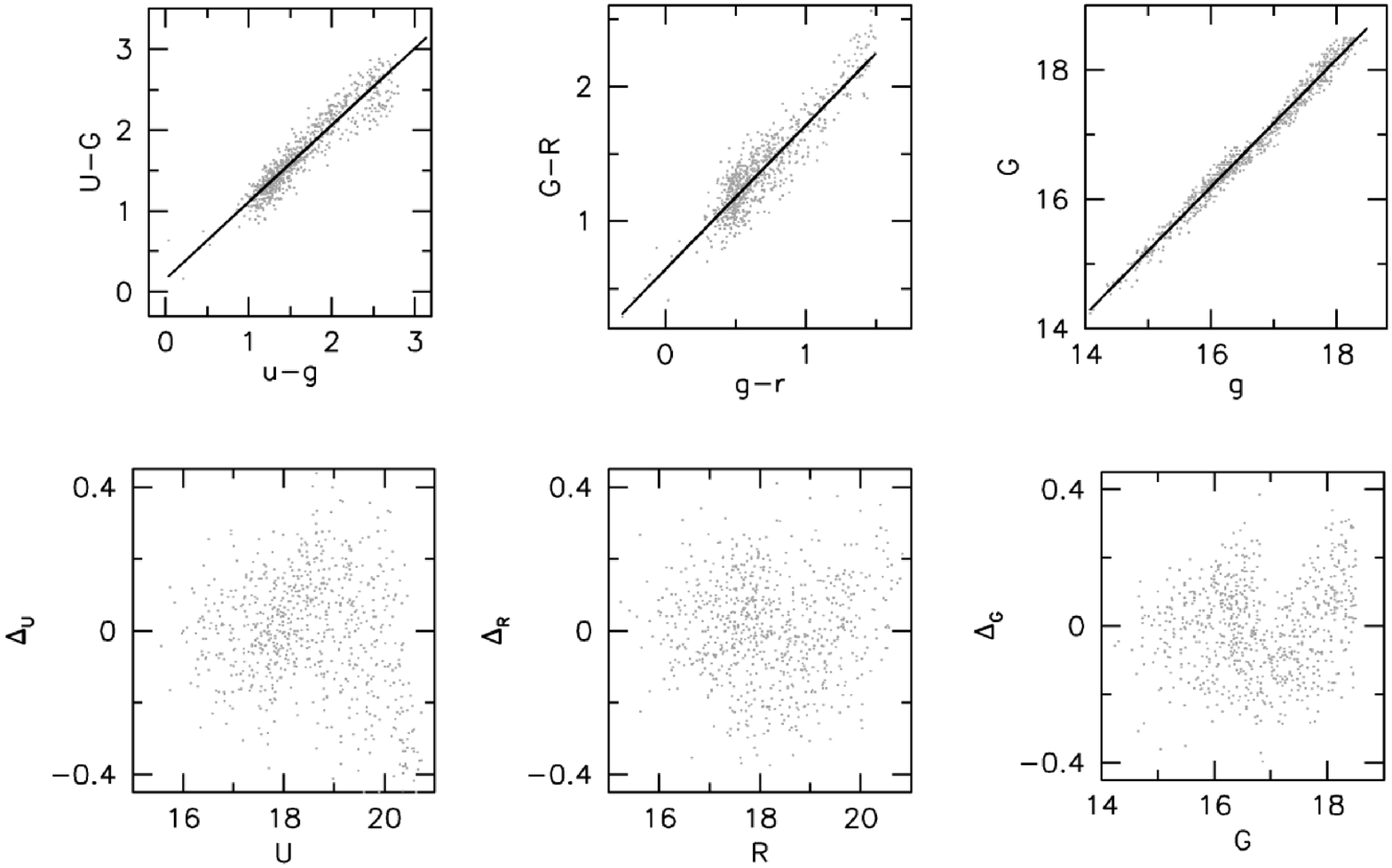}
\caption{The transformations between SDSS \emph{ugr} and
Becker's \emph{RGU} system in the upper row. In the lower three panels the deviations of the measured magnitude values from the calculated values is shown.}
 \label{rgu.UG} 
\end{figure*}

In Fig.~\ref{rgu.UG} in the upper three panels the distribution of the stars in the color planes
is shown, and the calculated transformations are plotted as straight
lines. Given the larger scatter, it is not surprising that the 
uncertainties of the coefficients are now
larger than for the \emph{UBVRI} transformations. Moreover, the stellar
data exhibit some wiggles, which we attribute to a nonlinear response
of the photographic plates used in the Basel survey.  Especially for
the $(u-g)$ transformation the nonlinearity appears to be a problem.
We investigated whether we could improve the transformation relations
by making them dependent on two color terms instead of one, but this
yielded no noticeable improvement.

\section{Discussion}

We presented empirical color transformations for the conversion of the
SDSS \emph{ugriz} photometric system into the Johnson-Cousins
\emph{UBVRI} system and into Becker's \emph{RGU} system, respectively.  These
are the first transformations between the SDSS and the Becker system,
whereas several earlier transformations between variants of the SDSS
and of the Johnson-Cousins system have been published.  For the Johnson-Cousins 
conversions, the novelty of our approach lies in the use of actual
SDSS data obtained with the 2.5-m SDSS telescope at Apache Point
Observatory and the use of well-defined Johnson-Cousins standard stars
taken from the lists of \citet{Landolt1992} and \citet{Stetson2000}.
For all transformations linear relations were found to be sufficient,
with a slope change in transformations involving $(V-R)$. 

A comparison with previously published transformations shows that they
qualitatively reproduce our transformations, but that they show systematic
differences that may amount to $\sim 0.1$ mag. This is due to the fact
that the earlier transformations were either done with data that used
other filter--detector--telescope combinations than that of the actual
2.5-m SDSS telescope and therefore differ intrinsically from the
actual SDSS data, or that the data are based on early versions of the
SDSS photometry catalogs before the SDSS photometry was in its final
form, or that non-standard Johnson-Cousins data were used.

For conversions between the SDSS and the Johnson-Cousins systems, we
tested the transformations for a possible metallicity dependence.
Some of the Stetson standard stars lie in a field centered on the
Draco dwarf spheroidal galaxy or on metal-poor ($\sim -2.0$ dex)
globular clusters.  Most of the stars in these fields are Population
II stars and have much lower metallicities than the majority of the
stars in Landolt's equatorial standard fields, where Population I
stars dominate.  The latter sample necessarily comprises stars with a
range of ages and abundances, while our Population II sample is
essentially limited to very old and very metal-poor stars.  The
calculated coefficients for the metal-poor stars result in slightly
different slopes as compared to the more metal-rich stars, which
affect very blue and very red stars most.  Although our metallicity
differentiation is quite crude, there does seem to be a slight
dependence of the transformations on the metallicity.  A more accurate
separation into metal-rich and metal-poor stars -- ideally aided by
spectroscopically measured stellar abundances -- is needed
to fully evaluate the magnitude of this trend. 

\begin{acknowledgements} 

We acknowledge support by the Swiss National Science Foundation
through grant 200020-105260 and 200020-105535.

Funding for the SDSS and SDSS-II has been provided by the Alfred P.\
Sloan Foundation, the Participating Institutions, the National Science
Foundation, the U.S.\ Department of Energy, the National Aeronautics
and Space Administration, the Japanese Monbukagakusho, the Max Planck
Society, and the Higher Education Funding Council for England. The
SDSS Web Site is http://www.sdss.org/.

The SDSS is managed by the Astrophysical Research Consortium for the
Participating Institutions. The Participating Institutions are the
American Museum of Natural History, Astrophysical Institute Potsdam,
University of Basel, Cambridge University, Case Western Reserve
University, University of Chicago, Drexel University, Fermilab, the
Institute for Advanced Study, the Japan Participation Group, Johns
Hopkins University, the Joint Institute for Nuclear Astrophysics, the
Kavli Institute for Particle Astrophysics and Cosmology, the Korean
Scientist Group, the Chinese Academy of Sciences (LAMOST), Los Alamos
National Laboratory, the Max-Planck-Institute for Astronomy (MPIA),
the Max-Planck-Institute for Astrophysics (MPA), New Mexico State
University, Ohio State University, University of Pittsburgh,
University of Portsmouth, Princeton University, the United States
Naval Observatory, and the University of Washington.

This research has made use of NASA's Astrophysics Data System
Bibliographic Services.
\end{acknowledgements}

\end{document}